\documentclass[preprints,article,accept,oneauthor,pdftex]{Definitions/mdpi}
\firstpage{1}
\pubvolume{xx}
\issuenum{1}
\articlenumber{5}
\pubyear{2019}
\copyrightyear{2019}
\history{Received: date; Accepted: date; Published: date}
\pdfoutput=1

\usepackage{fdsymbol}
\newcommand{\dd}{\mathrm{d}}
\newcommand{\DD}{\mathrm{D}}
\newcommand{\Lie}{\mathcal{L}}

\Title{Disformal Transformations in Scalar-Torsion Gravity}

\Author{Manuel Hohmann $^1$}

\AuthorNames{Manuel Hohmann}

\address[1]{%
$^{1}$ \quad Laboratory of Theoretical Physics, Institute of Physics, University of Tartu, W. Ostwaldi 1, 50411 Tartu, Estonia; manuel.hohmann@ut.ee}

\corres{Correspondence: manuel.hohmann@ut.ee}

\abstract{We study disformal transformations in the context of scalar extensions to teleparallel gravity, in which the gravitational interaction is mediated by the torsion of a flat, metric compatible connection. We find a generic class of scalar-torsion actions which is invariant under disformal transformations, and which possesses different invariant subclasses. For the most simple of these subclasses we explicitly derive all terms that may appear in the action. We propose to study actions from this class as possible teleparallel analogues of healthy beyond Horndeski theories.}

\keyword{teleparallel gravity; scalar-torsion gravity; disformal transformations}

\begin{document}
%%%%%%%%%%%%%%%%%%%%%%%%%%%%%%%%%%%%%%%%%%

%%%%%%%%%%%%%%%%%%%%%%%%%%%%%%%%%%%%%%%%%%
\section{Introduction}\label{sec:intro}
Scalar-tensor models form one of the largest and most well-studied classes of gravity theories. While this term most often refers to scalar-curvature theories~\cite{Fujii:2003pa}, which can be regarded as scalar extensions of general relativity based on its standard formulation in terms of curvature, it can also appropriately be applied to scalar field extensions of teleparallel gravity based on torsion~\cite{Hohmann:2018rwf,Hohmann:2018vle,Hohmann:2018dqh,Hohmann:2018ijr} or symmetric teleparallel gravity based on non-metricity~\cite{Jarv:2018bgs,Runkla:2018xrv}, and hence to scalar extensions of each of the three geometric pictures of gravity~\cite{BeltranJimenez:2019tjy}.

An interesting property of the aforementioned scalar-tensor theories is the possibility to perform scalar field dependent scale transformations of the fundamental fields defining the geometry, which may include a tetrad, a metric and an affine connection, depending on the particular class of theories under consideration. While in the most general class of metric-affine theories the metric and connection may be transformed independently, leading to different notions of invariance under such transformations~\cite{Iosifidis:2018zwo}, assuming a more specific geometry limits the possible transformations. In the scalar-curvature class, where the affine connection is fixed as the curvature-free, metric compatible Levi-Civita connection, this leads to the well-known possibility of conformal transformations~\cite{Flanagan:2004bz}, or the more general class of disformal transformations~\cite{Bekenstein:1992pj,Bettoni:2013diz} and its extensions~\cite{Zumalacarregui:2013pma,Ezquiaga:2017ner}. The latter are of particular interest, as they connect classes of gravity theories with second order field equations, such as the well-known Horndeski class~\cite{Horndeski:1974wa,Deffayet:2011gz,Kobayashi:2011nu}, to such higher derivative order theories which are healthy in the sense that they avoid Ostrogradsky instabilities due to the presence of constraints arising from degeneracies in their Lagrangians~\cite{Motohashi:2014opa,Langlois:2015cwa,Motohashi:2016ftl,BenAchour:2016fzp}. Theories in this larger class are known as beyond Horndeski models~\cite{Zumalacarregui:2013pma,Gleyzes:2014dya,Gao:2014soa,Langlois:2015cwa,Crisostomi:2016czh,BenAchour:2016fzp,Kobayashi:2019hrl}.

Also assuming the teleparallel geometry, which is based on a flat, metric compatible connection, allows for conformal transformations, which have been studied in the context of different theories~\cite{Yang:2010ji,Wright:2016ayu,Maluf:2011kf,Hohmann:2018vle,Hohmann:2018dqh,Hohmann:2018ijr}. In the present article we aim to generalize these previous studies by considering disformal transformations. In particular, we aim to construct a class of scalar-torsion theories of gravity which is closed under disformal transformations, and from which teleparallel analogues of Horndeski and beyond Horndeski models may be derived. Once such a class is identified, the physical properties of the contained theories may be studied in future work.

The outline of this article is as follows. In section~\ref{sec:geometry} we briefly review the geometric notions which are relevant for constructing scalar-torsion gravity theories. We study their behavior under disformal transformations and redefinitions of the scalar field in section~\ref{sec:disformal}. We apply these transformation rules in section~\ref{sec:invariant} in order to construct a disformally invariant class of scalar-torsion actions. We restrict this class to a simple subclass in section~\ref{sec:quadratic} and explicitly construct its action. We end with a conclusion in section~\ref{sec:conclusion}.

\section{Geometric notions}\label{sec:geometry}
We start our discussion of disformal transformations in scalar-torsion theories of gravity with a brief review of the geometric notions present in these theories and introducing a few convenient shorthand notations. Throughout this article we use the language of differential forms. In our convention lowercase Latin letters \(a, b, \ldots\) denote Lorentz indices, while we use lowercase Greek letters \(\mu, \nu, \ldots\) for spacetime indices; both types of indices take values from 0 to 3. Lorentz indices are raised and lowered using the Minkowski metric \(\eta_{ab}\), which we define with signature \((-,+,+,+)\).

The dynamical fields we consider in the covariant formulation of scalar-torsion gravity~\cite{Hohmann:2018rwf} are a tetrad \(\theta^a = \theta^a{}_{\mu}\dd x^{\mu}\), a flat Lorentz spin connection \(\omega^a{}_b = \omega^a{}_{b\mu}\dd x^{\mu}\), hence satisfying
\begin{equation}\label{eqn:spinlorentz}
\omega_{(ab)} = \frac{1}{2}(\eta_{ac}\omega^c{}_b + \eta_{bc}\omega^c{}_a) = 0
\end{equation}
and
\begin{equation}\label{eqn:spinflat}
R^a{}_b = \dd\omega^a{}_b + \omega^a{}_c \wedge \omega^c{}_b = 0\,,
\end{equation}
and a scalar field \(\phi\). A number of dependent quantities are derived from these fundamental fields. From the tetrad we define the dual tetrad \(e_a = e_a{}^{\mu}\partial_{\mu}\) such that
\begin{equation}
e_a \intprod \theta^b = e_a{}^{\mu}\theta^b{}_{\mu} = \delta_a^b\,,
\end{equation}
as well as the metric tensor
\begin{equation}
g = \eta_{ab}\theta^a \otimes \theta^b = \eta_{ab}\theta^a{}_{\mu}\theta^b{}_{\nu}\dd x^{\mu} \otimes \dd x^{\nu}\,.
\end{equation}
They are used to define the so-called musical isomorphisms: for a vector field \(v\) and a one-form \(\sigma\) we define
\begin{subequations}
\begin{align}
\sigma^{\sharp} &= (e_a \intprod \sigma)e^a = \eta^{ab}e_a{}^{\nu}\sigma_{\nu}e_b^{\mu}\partial_{\mu} = g^{\mu\nu}\sigma_{\nu}\partial_{\mu}\,,\\
\quad v^{\flat} &= (v \intprod \theta_a)\theta^a = \eta_{ab}v^{\nu}\theta^a{}_{\nu}\theta^b{}_{\mu}\dd x^{\mu} = g_{\mu\nu}v^{\nu}\dd x^{\mu}\,.
\end{align}
\end{subequations}
For the volume form of the tetrad we use the normalization
\begin{equation}\label{eqn:volume}
\mathrm{vol}_{\theta} = \frac{1}{4!}\epsilon_{abcd}\theta^a \wedge \theta^b \wedge \theta^c \wedge \theta^d = \theta^0 \wedge \theta^1 \wedge \theta^2 \wedge \theta^3\,.
\end{equation}
Using the exterior covariant derivative \(\DD\) of the spin connection \(\omega^a{}_b\) we further define the torsion
\begin{equation}\label{eqn:torsion}
T^a = \DD\theta^a = \dd\theta^a + \omega^a{}_b \wedge \theta^b\,.
\end{equation}
Note that from the torsion and the tetrad one may recover the spin connection as
\begin{equation}\label{eqn:invtorsion}
\omega_{ab} = \frac{1}{2}\left(e_b \intprod e_c \intprod T_a + e_c \intprod e_a \intprod T_b - e_a \intprod e_b \intprod T_c - e_b \intprod e_c \intprod \dd\theta_a - e_c \intprod e_a \intprod \dd\theta_b + e_a \intprod e_b \intprod \dd\theta_c\right)\theta^c\,.
\end{equation}
We also introduce a few helpful notations involving the derivatives of the scalar field, in a similar fashion as in~\cite{Ezquiaga:2016nqo}. For its Lie derivative with respect to the inverse tetrad we use the abbreviation
\begin{equation}
\phi_{,a} = \Lie_{e_a}\phi = e_a \intprod \dd\phi = e_a{}^{\mu}\partial_{\mu}\phi\,.
\end{equation}
We use this to define the one-forms
\begin{equation}
\psi_a = \phi_{,a}\dd\phi \quad \text{and} \quad \pi_a = \DD\phi_{,a} = \dd\phi_{,a} - \omega^b{}_a \wedge \phi_{,b}\,.
\end{equation}
Finally, we define the scalar
\begin{equation}
X = -\frac{1}{2}e_a \intprod \psi^a = -\frac{1}{2}\eta^{ab}\phi_{,a}\phi_{,b}\,,
\end{equation}
which represents the kinetic energy of the scalar field.

A number of helpful relations can be obtained for the quantities we introduced above. First note that
\begin{equation}
\psi_a \wedge \psi_b = 0\,, \quad \psi_a \wedge \dd\phi = 0\,, \quad \psi_a \wedge \theta^a = 0 \quad \text{and} \quad \dd\phi = \phi_{,a}\theta^a\,,
\end{equation}
which directly follows from the identity \(\dd\phi \wedge \dd\phi = 0\). For the one-forms we find the exterior covariant derivatives
\begin{equation}
\DD\psi_a = \pi_a \wedge \dd\phi \quad \text{and} \quad \DD\pi_a = 0\,.
\end{equation}
The derivative of the scalar field kinetic term is given by
\begin{equation}
\dd X = \DD X = -\eta^{ab}\phi_{,a}\DD\phi_{,b} = -\phi_{,a}\pi^a\,.
\end{equation}
We will make frequent use of these relations during the remainder of this article.

\section{Disformal transformations}\label{sec:disformal}
We now discuss the behavior of the terms introduced in the preceding section under disformal transformations and redefinitions of the scalar field, following closely the definitions used in~\cite{Ezquiaga:2017ner}. For this purpose we write the disformal transformation of the tetrad in the form
\begin{equation}
\bar{\theta}^a = \mathfrak{C}\theta^a + \mathfrak{D}\psi^a\,,
\end{equation}
where \(\mathfrak{C} = \mathfrak{C}(\phi, X)\) and \(\mathfrak{D} = \mathfrak{D}(\phi, X)\) are functions of the scalar field and its kinetic term.\footnote{Note that the disformal transformations discussed here are different from the extended conformal transformations discussed in~\cite{Capozziello:2007vj}; while the latter are always linear in the tetrad, this does not hold for the inhomogeneous part $\mathfrak{D}\psi^a$ used here.} In order to yield an invertible transformation, these functions must satisfy \(\mathfrak{C} \neq 0\) and \(\mathfrak{C} - 2X\mathfrak{D} \neq 0\). This can also be seen from the transformation behavior of the inverse tetrad, which is given by
\begin{equation}
\bar{e}_a = \frac{1}{\mathfrak{C}}\left(e_a - \frac{\mathfrak{D}}{\mathfrak{C} - 2X\mathfrak{D}}\psi^{\sharp}_a\right) = \frac{1}{\mathfrak{C}}\left(\delta_a^b - \frac{\mathfrak{D}}{\mathfrak{C} - 2X\mathfrak{D}}\phi_{,a}\phi_{,c}\eta^{bc}\right)e_b\,.
\end{equation}
For the scalar field we use a redefinition of the form
\begin{equation}
\bar{\phi} = f(\phi)
\end{equation}
with an invertible function \(f\). Finally, we leave the spin connection unchanged, \(\bar{\omega}^a{}_b = \omega^a{}_b\), since any transformation we would apply and which preserves the antisymmetry~\eqref{eqn:spinlorentz} and the flatness~\eqref{eqn:spinflat} of the connection could simply be absorbed into a local Lorentz transformation.

From the transformation laws of the basic quantities introduced above we can now derive the transformations of the dependent terms. For the metric we find the transformation
\begin{equation}
\bar{g} = \mathfrak{C}^2g + 2\mathfrak{D}(\mathfrak{C} - X\mathfrak{D})\dd\phi \otimes \dd\phi\,,
\end{equation}
while the volume form transforms as
\begin{equation}
\overline{\mathrm{vol}_{\theta}} = \mathfrak{C}^3(\mathfrak{C} - 2X\mathfrak{D})\mathrm{vol}_{\theta}\,.
\end{equation}
For the torsion we find
\begin{equation}\label{eqn:torsiontrans}
\bar{T}^a = \mathfrak{C}T^a + \mathfrak{C}_{,\phi}\dd\phi \wedge \theta^a + \mathfrak{C}_{,X}\dd X \wedge \theta^a + \mathfrak{D}\pi^a \wedge \dd\phi + \mathfrak{D}_{,X}\dd X \wedge \psi^a\,,
\end{equation}
where commas denote derivatives with respect to the corresponding function arguments. We then come to the transformation rules for the scalar field terms. First note that
\begin{equation}
\bar{\phi}_{,a} = \frac{f'}{\mathfrak{C} - 2X\mathfrak{D}}\phi_{,a}\,,
\end{equation}
from which follows the similarly simple transformation law
\begin{equation}
\bar{X} = \left(\frac{f'}{\mathfrak{C} - 2X\mathfrak{D}}\right)^2X\,.
\end{equation}
For the one-forms we introduced we find the transformation behavior
\begin{equation}
\bar{\psi}_a = \frac{f'^2}{\mathfrak{C} - 2X\mathfrak{D}}\psi_a\,,
\end{equation}
as well as
\begin{equation}
\bar{\pi}_a = \frac{f'}{\mathfrak{C} - 2X\mathfrak{D}}\pi_a + \left(\frac{f'}{\mathfrak{C} - 2X\mathfrak{D}}\right)_{,\phi}\psi_a + \left(\frac{f'}{\mathfrak{C} - 2X\mathfrak{D}}\right)_{,X}\phi_{,a}\dd X\,.
\end{equation}
With these transformation rules at hand we can now study how any action which is constructed from the fundamental fields will behave under disformal transformations. This will be done in the following sections.

\section{A disformally invariant class of theories}\label{sec:invariant}
We now apply the general transformation laws of geometric quantities derived in the preceding section to a class of actions composed from these quantities. In particular, we aim to construct a class of scalar-torsion theories of gravity, which
\begin{enumerate}
\item
contains new general relativity~\cite{Hayashi:1979qx} as a sub-class,
\item
is closed under disformal transformations,
\item
whose action is given by a sum \(\sum F_k(\phi, X)Q_k\) of terms \(Q_k\) constructed from the objects discussed in section~\ref{sec:geometry}, with coefficients given by free functions \(F_k\) of the scalar field \(\phi\) and its kinetic energy \(X\).
\end{enumerate}
While this choice of requirements may seem arbitrary at first glance, it turns out that it leads to a class with various interesting and well-studied examples, as we will see below. In order to construct this class of actions, we consider the terms
\begin{equation}
A^{Ia}{}_{bc} = e_c \intprod e_b \intprod A^{Ia}, \quad I = 1, \ldots, 7\,,
\end{equation}
where the two-forms \(A^{Ia}\) are given by
\begin{gather}
A^{1a} = T^a\,, \quad
A^{2a} = \dd\phi \wedge (\dd\phi^{\sharp} \intprod T^a)\,, \quad
A^{3a} = \dd\phi \wedge \theta^a\,, \quad
A^{4a} = \dd X \wedge \theta^a\,, \nonumber\\
A^{5a} = \dd X \wedge \psi^a\,, \quad
A^{6a} = \pi^a \wedge \dd\phi\,, \quad
A^{7a} = A^{3a}\eta^{bc}\phi_{,b}X_{,c}\,.\label{eqn:actionterms}
\end{gather}
Their coefficients \(A^{Ia}{}_{bc}\) in the tetrad basis take the explicit forms
\begin{gather}
A^{1a}{}_{bc} = T^a{}_{bc}\,, \quad
A^{2a}{}_{bc} = 2T^{ad}{}_{[c}\phi_{,b]}\phi_{,d}\,, \quad
A^{3a}{}_{bc} = 2\phi_{,[b}\delta^a_{c]}\,, \quad
A^{4a}{}_{bc} = 2X_{,[b}\delta^a_{c]}\,, \nonumber\\
A^{5a}{}_{bc} = 2\eta^{ad}X_{,[b}\phi_{,c]}\phi_{,d}\,, \quad
A^{6a}{}_{bc} = 2\phi_{,[c}e_{b]} \intprod \pi^a\,, \quad
A^{7a}{}_{bc} = 2\eta^{de}\phi_{,d}X_{,e}\phi_{,[b}\delta^a_{c]}\,.
\end{gather}
We have chosen this particular set of two-forms because they contain the torsion \(T^a\), which is a crucial ingredient for building any teleparallel gravity action, and because it is the minimal set of two-forms such that the conformally transformed quantities \(\bar{A}^{Ia}{}_{bc}\) can be expressed in the form
\begin{equation}\label{eqn:acttermtrans}
\bar{A}^{Ia}{}_{bc} = \bar{e}_c \intprod \bar{e}_b \intprod \bar{A}^{Ia} = \sum_{J = 1}^7M^I{}_J(\phi, X)A^{Ja}{}_{bc}
\end{equation}
in terms of the original quantities \(A^{Ia}{}_{bc}\), with coefficients \(M^I{}_J\) which are functions of the scalar field and its kinetic term only. This can most easily be seen by successively applying a general disformal transformation to the torsion term \(A^{1a}{}_{bc}\) and all terms which result from this operation, which yields explicit expressions for the remaining terms \(A^{Ia}{}_{bc}\) and the coefficients \(M^I{}_J\). In particular, from the transformation~\eqref{eqn:torsiontrans} of the torsion we find the terms
\begin{equation}
M^1{}_1 = \frac{1}{\mathfrak{C}}\,, \quad\!
M^1{}_6 = -M^1{}_2 = \frac{\mathfrak{D}}{\mathfrak{C}\mathfrak{E}}\,, \quad\!
M^1{}_3 = \frac{\mathfrak{C}_{,\phi}}{\mathfrak{C}\mathfrak{E}}\,, \quad\!
M^1{}_4 = \frac{\mathfrak{C}_{,X}}{\mathfrak{C}^2}\,, \quad\!
M^1{}_5 = \frac{\mathfrak{C}\mathfrak{D}_{,X} - \mathfrak{D}\mathfrak{C}_{,X}}{\mathfrak{C}^2\mathfrak{E}}\,, \quad\!
M^1{}_7 = -\frac{\mathfrak{D}\mathfrak{C}_X}{\mathfrak{C}^2\mathfrak{E}}\,,
\end{equation}
where we introduced the abbreviation \(\mathfrak{E} = \mathfrak{C} - 2X\mathfrak{D}\). Observe that we find non-vanishing coefficients for all terms~\eqref{eqn:actionterms}, which shows that they indeed form a minimal set. The transformation~\eqref{eqn:torsiontrans} of the torsion further yields the coefficients
\begin{equation}
M^2{}_2 = \frac{f'^2}{\mathfrak{E}^3}\,, \quad
M^2{}_3 = -\frac{2Xf'^2\mathfrak{C}_{,\phi}}{\mathfrak{C}\mathfrak{E}^3}\,, \quad
M^2{}_5 = \frac{f'^2(\mathfrak{C}_{,X} - 2X\mathfrak{D}_{,X})}{\mathfrak{C}\mathfrak{E}^3}\,, \quad
M^2{}_6 = -\frac{2Xf'^2\mathfrak{D}}{\mathfrak{C}\mathfrak{E}^3}\,, \quad
M^2{}_7 = \frac{f'^2\mathfrak{C}_{,X}}{\mathfrak{C}\mathfrak{E}^3}\,.
\end{equation}
For the scalar field terms we find that they transform into each other, and partially even into themselves, up to a factor, so that no further torsion terms arise. The latter holds in particular for \(A^{3a}\) and \(A^{5a}\), where we obtain
\begin{equation}
M^3{}_3 = \frac{f'}{\mathfrak{E}}\,, \quad
M^5{}_5 = \frac{f'^4\mathfrak{F}}{\mathfrak{C}\mathfrak{E}^5}\,,
\end{equation}
with the abbreviation \(\mathfrak{F} = \mathfrak{C} + 2X\mathfrak{D} - 2X\mathfrak{C}_{,X} + 4X^2\mathfrak{D}_{,X}\). A similarly simple transformation is found for \(A^{6a}\), which satisfies
\begin{equation}
M^6{}_5 = \frac{f'^2(2\mathfrak{D} - \mathfrak{C}_{,X} + 2X\mathfrak{D}_{,X})}{\mathfrak{C}\mathfrak{E}^3}\,, \quad
M^6{}_6 = \frac{f'^2}{\mathfrak{C}\mathfrak{E}^2}\,.
\end{equation}
Finally, for the remaining two terms we have the transformation
\begin{equation}
M^4{}_3 = \frac{2Xf'\mathfrak{G}}{\mathfrak{E}^2}\,, \quad
M^4{}_4 = \frac{f'^2\mathfrak{F}}{\mathfrak{C}\mathfrak{E}^3}\,, \quad
M^4{}_7 = -\frac{f'^2\mathfrak{D}\mathfrak{F}}{\mathfrak{C}\mathfrak{E}^4}\,, \quad
M^7{}_3 = -\frac{4X^2f'^3\mathfrak{G}}{\mathfrak{E}^4}\,, \quad
M^7{}_7 = \frac{f'^4\mathfrak{F}}{\mathfrak{E}^6}\,,
\end{equation}
with the abbreviation
\begin{equation}
\mathfrak{G} = \left(\frac{f'}{\mathfrak{E}}\right)_{,\phi} = \frac{f''\mathfrak{E} - f'\mathfrak{E}_{,\phi}}{\mathfrak{E}^2} = \frac{f''}{\mathfrak{C} - 2X\mathfrak{D}} - \frac{f'(\mathfrak{C}_{,\phi} - 2X\mathfrak{D}_{,\phi})}{(\mathfrak{C} - 2X\mathfrak{D})^2}\,.
\end{equation}
We thus find that the terms we chose indeed form a set which is invariant under disformal transformations.

From the transformation behavior of the action terms discussed above follows that any expression given by a sum of products and contractions of these terms (using the Minkowski metric to raise, lower and contract Lorentz indices), with functions of the scalar field and its kinetic term as coefficients, again transforms into an expression of the same form. Note that such an expression is necessarily a scalar if all Lorentz indices are contracted. Together with the volume form~\eqref{eqn:volume}, which likewise transforms into itself up to a factor, we may thus construct a generic gravitational action of the form
\begin{equation}\label{eqn:disinvaction}
S_g = \int_M\mathrm{vol}_{\theta}\sum_N\sum_{I_1} \cdots \sum_{I_N}H_{I_1 \cdots I_Na_1 \cdots a_N}{}^{b_1 \cdots b_Nc_1 \cdots c_N}A^{I_1a_1}{}_{b_1c_1} \cdots A^{I_Na_N}{}_{b_Nc_N}\,,
\end{equation}
where the coefficients \(H_{I_1 \cdots I_Na_1 \cdots a_N}{}^{b_1 \cdots b_Nc_1 \cdots c_N}\) are composed from Minkowski metrics, Kronecker symbols and functions of the scalar field and its kinetic term; the latter implies that the number of Lorentz indices will be even, so that also \(N\) must be even. Applying a disformal transformation to this generic action preserves its form, up to replacing the coefficients \(H\) with transformed coefficients \(\bar{H}\), which can be obtained using the transformation rules~\eqref{eqn:acttermtrans}. Complementing this gravitational action with a suitable matter action, which exhibits the same type of invariance, hence leads to a disformally invariant class of scalar-torsion theories.

Note that the theories defined by the action~\eqref{eqn:disinvaction} in general possess field equations of higher than second derivative order, and thus potentially suffer from Ostrogradsky instabilities. However, one may expect that theories which are disformally equivalent to second order theories avoid this problem and may thus be considered healthy, as it is also the case for scalar-curvature theories~\cite{Zumalacarregui:2013pma}, and that the class we constructed here even contains the teleparallel formulation of a number of beyond Horndeski models~\cite{Gleyzes:2014dya,Gao:2014soa,Langlois:2015cwa,Crisostomi:2016czh,BenAchour:2016fzp,Kobayashi:2019hrl}. Further studies are required to show whether this is indeed the case.

We finally remark that any terms for a given order \(N\) in the action~\eqref{eqn:disinvaction} will transform only into terms with the same or lower order \(\bar{N} \leq N\). One may thus restrict this class of theories to any finite maximum order. A case of particular interest is given by at most quadratic order, and will be discussed in the following section.

\section{The quadratic class of actions}\label{sec:quadratic}
While the action~\eqref{eqn:disinvaction} constructed in the previous section is still very generic, it contains a number of simpler cases, which can be obtained by restricting the order \(N\) to a fixed maximum value. In this section we consider the at most quadratic case \(N \leq 2\), which allows to write the coefficients as \(H\) (without any indices) for the zeroth order \(N = 0\) and
\begin{equation}
H_{I_1I_2a_1a_2}{}^{b_1b_2c_1c_2} = U_{I_1I_2}\eta_{a_1a_2}\eta^{b_1b_2}\eta^{c_1c_2} + V_{I_1I_2}\delta_{a_2}^{[c_1}\eta^{b_1][b_2}\delta^{c_2]}_{a_1} + W_{I_1I_2}\delta_{a_1}^{[c_1}\eta^{b_1][b_2}\delta^{c_2]}_{a_2}
\end{equation}
for the second order \(N = 2\), with free functions \(H, U_{I_1I_2}, V_{I_1I_2}, W_{I_1I_2}\) of the scalar field and its kinetic terms. One easily checks that any term at order \(N = 2\) must be of this form, since there is no other possibility to form a scalar out of the terms \(A^{Ia}{}_{bc}\) due to the antisymmetry in their lower two indices. In order to further study this class of theories, it is helpful to explicitly calculate the appearing terms, which are given by the following exhaustive list:

\begin{itemize}
\item
Terms quadratic in the torsion only:
\begin{equation}
Q_1 = T^{abc}T_{abc}\,, \quad
Q_2 = T^{abc}T_{cba}\,, \quad
Q_3 = T^a{}_{ac}T_b{}^{bc}\,.
\end{equation}
These are the terms which form the action of new general relativity~\cite{Hayashi:1979qx}, and which arise from considering quadratic combinations of \(A^{1a}{}_{bc}\). Note, however, that here we consider these terms not with constant coefficients, but with coefficients which are functions of the scalar field and its kinetic term.
\item
Terms quadratic in the torsion which contain the scalar field:
\begin{gather}
Q_4 = T^{abe}T^{cd}{}_e\phi_{,a}\phi_{,b}\phi_{,c}\phi_{,d}\,, \quad
Q_5 = T^{abc}T^d{}_{dc}\phi_{,a}\phi_{,b}\,, \quad
Q_6 = T^{cda}T_{cd}{}^b\phi_{,a}\phi_{,b}\,, \nonumber\\
Q_7 = T^{cda}T_{dc}{}^b\phi_{,a}\phi_{,b}\,, \quad
Q_8 = T^{cda}T^b{}_{cd}\phi_{,a}\phi_{,b}\,, \quad
Q_9 = T^{acd}T^b{}_{cd}\phi_{,a}\phi_{,b}\,.
\end{gather}
These are all terms which can be formed from contractions of the square of the torsion tensor with derivatives of the scalar field. We find such terms by considering products among \(A^{2a}{}_{bc}\) and \(A^{6a}{}_{bc}\), or of one of them with \(A^{1a}{}_{bc}\). The appearance of the last term is not obvious from a naive calculation of the action terms~\eqref{eqn:disinvaction} at order \(N = 2\); this will be explained below.
\item
Terms linear in the torsion:
\begin{equation}
Q_{10} = T_a{}^{ab}\phi_{,b}\,, \quad
Q_{11} = T^{abc}e_a \intprod \pi_b\phi_{,c}\,, \quad
Q_{12} = T_a{}^{ab}X_{,b}\,, \quad
Q_{13} = T^{abc}\phi_{,a}\phi_{,b}X_{,c}\,.
\end{equation}
The first term is the familiar kinetic coupling to the vector torsion, which is present in various generalized scalar-torsion theories~\cite{Hohmann:2018dqh}, while the remaining terms contain also second order derivatives of the scalar field. We find them in particular by taking products of \(A^{1a}{}_{bc}\) or \(A^{2a}{}_{bc}\) with any of the other building blocks. Here we used the fact that
\begin{equation}
X_{,a} = -e_a \intprod \pi^b\phi_{,b}
\end{equation}
in order to find identical terms. Further, one may wonder why there are no terms \(T^{bac}e_a \intprod \pi_b\phi_{,c}\) and \(T^{cab}e_a \intprod \pi_b\phi_{,c}\) listed here. This is due to the fact that
\begin{equation}
e_a \intprod \pi_b - e_b \intprod \pi_a = T^c{}_{ba}\phi_{,c}\,,
\end{equation}
so that
\begin{equation}
T^{bac}e_a \intprod \pi_b\phi_{,c} = T^{abc}e_a \intprod \pi_b\phi_{,c} + T^{abc}T^d{}_{ab}\phi_{,c}\phi_{,d} = Q_8 + Q_{11}
\end{equation}
as well as
\begin{equation}
T^{cab}e_a \intprod \pi_b\phi_{,c} = -\frac{1}{2}T^{cab}T^d{}_{ab}\phi_{,c}\phi_{,d} = -\frac{1}{2}Q_9\,.
\end{equation}
The latter causes the non-obvious appearance of the term \(Q_9\) mentioned before.
\item
Terms involving the scalar field only:
\begin{equation}
Q_{14} = X_{,a}X^{,a}\,, \quad
Q_{15} = X_{,a}\phi^{,a}\,, \quad
Q_{16} = (e_a \intprod \pi^b)(e_b \intprod \pi^a)\,, \quad
Q_{17} = e_a \intprod \pi^a\,.
\end{equation}
These terms can be found by taking products of the terms \(A^{3a}{}_{bc}, \ldots, A^{7a}{}_{bc}\). Note that also terms proportional to \(X\) appear from taking the product of the term \(A^{3a}{}_{bc}\) with itself. We do not list these terms here, since they can be absorbed into the coefficient function \(H(\phi, X)\) at the zeroth order \(N = 0\), and thus do not yield any new terms. This leaves only the terms containing second order derivatives listed above.
\item
Products of the simple terms:
\begin{equation}
Q_{18} = Q_{15}Q_{17}\,, \quad
Q_{19} = Q_{10}Q_{17}\,, \quad
Q_{20} = Q_{10}Q_{15}\,, \quad
Q_{21} = Q_{17}^2\,, \quad
Q_{22} = Q_{15}^2\,, \quad
Q_{23} = Q_{10}^2\,.
\end{equation}
In addition to linear combinations the previously listed terms we also find a few terms which involve their products, and which must therefore be taken into account separately in order to write the final action again as a linear combination.
\end{itemize}

A detailed analysis shows that the action~\eqref{eqn:disinvaction} at the second order \(N = 2\) does not contain all possible linear combinations of the terms \(Q_1, \ldots, Q_{23}\) listed above. It turns out that the term \(Q_{23}\) appears only together with other terms, and may hence be omitted if one defines
\begin{gather}
\tilde{Q}_4 = Q_4 - 2XQ_{23}\,, \quad
\tilde{Q}_5 = Q_5 + Q_{23}\,, \quad
\tilde{Q}_7 = Q_7 - Q_{23}\,, \nonumber\\
\tilde{Q}_9 = Q_9 + 2Q_{23}\,, \quad
\tilde{Q}_{11} = Q_{11} + Q_{23}\,, \quad
\tilde{Q}_{16} = Q_{16} + Q_{23}\,,
\end{gather}
and \(\tilde{Q}_k = Q_k\) for all other terms. Further defining \(\tilde{Q}_0 = 1\), one may then write the most general action~\eqref{eqn:disinvaction} of order \(N \leq 2\) as
\begin{equation}
S_g = \int_M\mathrm{vol}_{\theta}\sum_{k = 0}^{22}F_k(\phi, X)\tilde{Q}_k\,,
\end{equation}
where \(F_0, \ldots, F_{22}\) are free functions of the scalar field and its kinetic term. It follows from the discussion detailed in the preceding section~\ref{sec:invariant} that this class of actions is invariant under disformal transformations, i.e., that any action of this form transforms into another action of the same form, but with different coefficient functions \(\bar{F}_k\) replacing the original functions \(F_k\), when a disformal transformation is applied. The form of these transformed functions can be derived from the transformation behavior~\eqref{eqn:acttermtrans} of the terms from which the action is constructed; we omit the result here for brevity.

The class of theories presented here contains a number of interesting examples, some of which have already been studied, such as an analogue of scalar-curvature gravity~\cite{Hohmann:2018ijr} and a scalar-torsion generalization of new general relativity, which includes conformal teleparallel gravity~\cite{Maluf:2011kf}. Note, however, that the full class of Horndeski models and its recently proposed teleparallel extension~\cite{Bahamonde:2019shr} are not included here, since the Horndeski Lagrangian \(\mathcal{L}_5\) requires terms which are cubic in second order derivatives of the scalar field, and which appear only in the next order \(N = 4\). A full treatment of these fourth order theories would exceed the scope of this article.

\section{Conclusion}\label{sec:conclusion}
We have discussed disformal transformations in the context of scalar-torsion extensions of teleparallel gravity and derived the transformation behavior of the most important geometric quantities used in these theories. Based on these transformation rules, we have constructed a class of scalar-torsion theories of gravity which is closed under disformal transformations, in the sense that applying a disformal transformation to any theory of this class yields a theory which again lies in the same class. The Lagrangian of these theories is given by a polynomial of arbitrary order in a number of tensorial building blocks. Further, we considered the restricted case of actions which are at most quadratic in these building blocks, and which by itself forms a disformally invariant subclass. For the latter class we explicitly derived all terms whose linear combination constitutes the general form of the action.

Our results show that the studies of disformal transformations in scalar-curvature gravity~\cite{Zumalacarregui:2013pma,Ezquiaga:2017ner} can be generalized also to scalar-torsion theories. In this work we have undertaken a first step in this direction, by focusing on general disformal transformations only, and by explicitly constructing an invariant class of actions by applying these transformations to the torsion tensor. One may follow a similar approach as in the scalar-curvature case and consider the restricted case of special disformal transformations, which do not depend on the kinetic term of the scalar field, or the even more general case of extended disformal transformations, which also include second order derivatives of the scalar field. Given these different sets of transformations, one may study their orbits in the space of scalar-torsion Lagrangians, and investigate whether these yield healthy higher order theories, in a similar fashion to the healthy beyond Horndeski theories~\cite{Zumalacarregui:2013pma,Gleyzes:2014dya,Gao:2014soa,Langlois:2015cwa,Crisostomi:2016czh,BenAchour:2016fzp,Kobayashi:2019hrl}. Another question of particular interest is whether the recently proposed teleparallel extension of Horndeski gravity~\cite{Bahamonde:2019shr} is invariant under special disformal transformations, as it is the case for the scalar-curvature Horndeski class~\cite{Bettoni:2013diz}.

A different direction of further studies besides exhaustively charting the landscape of scalar-torsion theories is to study the phenomenology of extended models such as those presented in this article. A promising approach for this work is given by extending the formalism of conformal invariants, which was originally developed for scalar-curvature theories of gravity~\cite{Jarv:2014hma} and later extended to scalar-torsion gravity~\cite{Hohmann:2018ijr}, to include also disformal transformations. Such an extended formalism would be useful for both scalar-curvature and scalar-torsion gravity, and potentially allow expressing observational properties of these theories in terms of invariant quantities. Potential applications include the post-Newtonian limit~\cite{Hohmann:2015kra} and cosmological dynamics, either following the Noether symmetry approach, as shown for scalar-curvature Horndeski gravity~\cite{Capozziello:2018gms}, or by using methods used in $f(T)$ cosmology for flat~\cite{Hohmann:2017jao} and non-flat~\cite{Capozziello:2018hly} spatial geometry.

%%%%%%%%%%%%%%%%%%%%%%%%%%%%%%%%%%%%%%%%%%
\vspace{6pt}

%%%%%%%%%%%%%%%%%%%%%%%%%%%%%%%%%%%%%%%%%%
%% optional
%\supplementary{The following are available online at \linksupplementary{s1}, Figure S1: title, Table S1: title, Video S1: title.}


%%%%%%%%%%%%%%%%%%%%%%%%%%%%%%%%%%%%%%%%%%
%\authorcontributions{For research articles with several authors, a short paragraph specifying their individual contributions must be provided. The following statements should be used ``conceptualization, X.X. and Y.Y.; methodology, X.X.; software, X.X.; validation, X.X., Y.Y. and Z.Z.; formal analysis, X.X.; investigation, X.X.; resources, X.X.; data curation, X.X.; writing--original draft preparation, X.X.; writing--review and editing, X.X.; visualization, X.X.; supervision, X.X.; project administration, X.X.; funding acquisition, Y.Y.'', please turn to the  \href{http://img.mdpi.org/data/contributor-role-instruction.pdf}{CRediT taxonomy} for the term explanation. Authorship must be limited to those who have contributed substantially to the work reported.}

%%%%%%%%%%%%%%%%%%%%%%%%%%%%%%%%%%%%%%%%%%
\funding{This work was supported by the Estonian Research Council through the Personal Research Funding project PRG356 and by the European Regional Development Fund through the Center of Excellence TK133 ``The Dark Side of the Universe''.}

%%%%%%%%%%%%%%%%%%%%%%%%%%%%%%%%%%%%%%%%%%
\acknowledgments{The author thanks the organizers of the conference Teleparallel Universes in Salamanca for the kind invitation, and Tomi Koivisto for helpful comments.}

%%%%%%%%%%%%%%%%%%%%%%%%%%%%%%%%%%%%%%%%%%
\conflictsofinterest{The author declares no conflict of interest.}


%=====================================
% References, variant A: internal bibliography
%=====================================

%=====================================
% References, variant B: external bibliography
%=====================================
\externalbibliography{yes}
\bibliography{scaltors}
\end{document}